\documentclass[prc,aps,twocolumn,groupedaddress,showpacs,floatfix]{revtex4}
\usepackage{graphicx}
\usepackage[usenames]{color}

\usepackage{amsbsy}

\newcommand{\beq}{\begin{equation}}
\newcommand{\eeq}{\end{equation}}
\newcommand{\bea}{\begin{eqnarray}}
\newcommand{\eea}{\end{eqnarray}}

\newcommand{\benn}{\begin{displaymath}}
\newcommand{\eenn}{\end{displaymath}}

\begin{document}

\title{Low-density neutron matter}

\author{ Alexandros Gezerlis$^{1,2,3}$ and J. Carlson$^1$ }
\affiliation{$^1$ Theoretical Division, Los Alamos National Laboratory, Los Alamos, New Mexico 87545,  USA}
\affiliation{$^2$ Department of Physics, University of Illinois at Urbana-Champaign, Urbana, Illinois 61801, USA}
\affiliation{$^3$ Department of Physics, University of Washington, Seattle, Washington, 98195, USA}

\begin{abstract}
The properties of low-density neutron matter are important for the understanding of neutron star crusts and the exterior of large neutron-rich nuclei.
We examine various properties of dilute neutron matter using quantum Monte Carlo methods, with $s$- and $p$-wave terms in the interaction. Our results provide a smooth evolution
of the equation of state and pairing gap from extremely small densities, where analytic expressions are available, up to the strongly interacting regime probed experimentally and described theoretically in cold atomic systems, where $k_F \approx 0.5~\mbox{fm}^{-1}$ and the pairing gap becomes of the order of magnitude of the Fermi energy. We also present results for the momentum distribution and pair distributions, displaying the same evolution from weak to strong coupling. Combined with previous quantum Monte Carlo and other calculations at moderate densities,
these results provide strong constraints on the neutron matter equation of state up to saturation densities.
\end{abstract}

\date{\today}


\maketitle

\section{Introduction}

The equation of state and the pairing gap of neutron matter at low densities are important for describing the properties of the inner crusts of
neutron stars and provide significant constraints for density-functional theories of large neutron-rich nuclei. Equation of state results at
larger densities ($ \rho \geq 0.04$ fm$^{-3}$) have been used for some time to constrain  Skyrme and other density functional approaches 
to large nuclei \cite{Brown:2000,Stone:2003}.
More recently, the density-dependence of the $^1\mbox{S}_0$ gap in low-density neutron matter has been used to constrain Skyrme-Hartree-Fock-Bogoliubov treatments and especially their description of neutron-rich nuclei \cite{Chamel:2008}.  At extremely low densities the equation of state and pairing gap can be expressed as analytically known functions of $(k_F a)$, the product of the
Fermi momentum and the neutron-neutron scattering length.  Our results provide a smooth connection between these
analytic results and previous calculations of neutron matter at larger densities, where the gaps become smaller and the superfluid
properties are less relevant to the equation of state.

The properties of low-density neutron matter are particularly important for describing the inner crust of a neutron
star, which is composed of a lattice of neutron-rich nuclei along with a gas of neutrons and electrons.  The neutron gas at low
densities is expected to be superfluid; the evolution of the equation of state and the pairing gap will impact the static and dynamic
properties of the inner crust of the neutron star.  A cold neutron star will have a temperature from
$10^6$ K to $10^9$ K  ($\sim 0.1$ keV to $0.1$ MeV); hence the low-density neutron gas is superfluid because the critical temperature is expected to be larger, approximately $10^{10}$ K ($\sim 1$ MeV). The most basic aspects of a superfluid gas embedded in a lattice of nuclei are described by zero-temperature infinite pure neutron matter.  Corrections to this picture include gradient terms in the density induced by the ionic lattice. While these corrections are also important for density-functional theories of nuclei, we leave their study to future investigations.

Superfluidity in neutron matter is often connected to cooling observations of neutron stars: the specific heat in a superfluid is exponentially suppressed, a fact which is consistent with observations of cooling quiescent neutron stars \cite{Brown:2009}.
Furthermore, in the presence of a neutron $^1\mbox{S}_0$ gap, the neutron-neutron bremsstrahlung reaction rate is also suppressed. 
Cooper-pair breaking/formation neutrino emission processes that occur near the transition temperature are also relevant to the cooling of neutron stars during the crust's thermal relaxation \cite{Steiner:2009, Page:2009}.  While many of these phenomena are not critically dependent upon the
magnitude of the gap,
recent heat-conduction mechanisms in magnetars require superfluid phonons and their interaction with the lattice ions \cite{Aguilera:2009}. These may be more sensitive to the magnitude of the gap.

The neutron matter equation of state \cite{Friedman:1981,Akmal:1998,Carlson:Morales:2003,Schwenk:2005,Margueron:2007b,Baldo:2008,Epelbaum:2008b,Epelbaum:2008a,Hebeler:2009,Rios:2009} and pairing gap \cite{Lombardo:2001,Dean:2003,Chen:1993,Wambach:1993,Schulze:1996,Schwenk:2003,Muether:2005,Fabrocini:2005,Cao:2006,Margueron:2008,Gandolfi:2008,Abe:2009,Gandolfi:2009} have been the subject of many studies over the years, 
with quite different results, particularly for the pairing gap.
Now, however, cold-atom experiments can 
mimic the properties of dilute neutron matter, giving nearly direct constraints on its properties.
In cold atoms the interaction can be tuned through Feshbach resonances to produce a  specific scattering 
length $a$, while the effective range $r_e$ between the atoms is nearly zero. In low-density neutron matter, on the other hand,
the particle-particle interaction has a scattering length which is
very large, $\approx -18.5$ fm, much larger than the typical separation between
neutrons.  The effective range is much smaller than the scattering length,
$r_e \approx 2.7$ fm, so $ |r_e / a| \approx 0.15$, but only at very low densities is the
effective range much smaller than the interparticle spacing.  We directly compare results in neutron matter and cold atoms to try 
to understand the impact of the effective range theoretically; it may also be possible to use narrow and wide resonances in cold atoms to study
this experimentally.\cite{Marcelis:2008}

In a recent article \cite{Gezerlis:2008}, we examined the similarities of cold atoms with neutron matter by calculating the $T=0$ equations of state and pairing gaps. The interaction used for the cold atoms had an infinitesimal range and the scattering length was varied to obtain results from $k_F a = -1$ to $-10$.
For the case of neutron matter, we took the $s$-wave part of the AV18 \cite{Wiringa:1995} interaction that fits $s$-wave nucleon-nucleon scattering
very well at both low- and high-energies. In this work, we extend our approach to include $p$-wave interactions,
examining their effects on the equation of state and superfluid pairing gap.  Additional corrections due to higher
partial waves and three-nucleon interactions are expected to be quite small in this density regime.
We also calculate additional properties of neutron matter, including the quasiparticle excitation spectrum, the momentum distribution, 
and pair-distribution functions.  

All calculations are performed using quantum Monte Carlo (QMC) techniques (section \ref{sec:qmc}), including Variational Monte Carlo (VMC) and Green's Function Monte Carlo (GFMC) methods. We compare our results to analytic calculations at very low densities (section \ref{sec:exact}), and to BCS calculations over the range of densities we consider (section \ref{sec:bcs}).  The BCS calculations are also used to try to understand and estimate the finite-size effects in the QMC simulations. Although the BCS results are not expected to be quantitatively accurate, they do provide a useful benchmark for comparisons and for understanding physical effects beyond the mean-field treatment of pairing.

\section{Analytic Results}
\subsection{Weak coupling}
\label{sec:exact} 
At extremely low densities ($| k_F a | << 1 $) the effective coupling between 
neutrons is weak and neutron matter properties can be calculated 
analytically.
The ground-state energy of normal (i.e. non-superfluid) matter 
in this regime was calculated by Lee and Yang in 1957: \cite{Lee:1957}
\begin{equation}
\frac{E}{E_{FG}}  =  1 + \frac{10}{9\pi}k_Fa + \frac{4}{21\pi^2} \left
( 11 - 2 \ln2 \right ) \left ( k_Fa \right )^2~,
\label{eq:Lee}
\end{equation}
where $E_{FG}$ is the energy of a free Fermi gas at the same density as the interacting gas.  While this expression ignores the contributions of superfluidity, these are exponentially small in (1/$k_F a$).  In the next section we
compare these results to QMC calculations for $|k_F a| \geq 1$.

The pairing gap at weak coupling is also known analytically. The
mean-field BCS approach described below [Eq. \ref{deltaconteq}]
reduces in this limit to:
\begin{equation}
\Delta^0_{BCS}(k_F) = \frac{8}{e^2} \frac{\hbar^2 k_F^2}{2m} \exp\left( \frac{\pi}{2ak_F}\right)~.
\label{eq:weakBCS}
\end{equation}
However, as was shown in 1961 by Gorkov and Melik-Barkhudarov,\cite{Gorkov:1961} the BCS result acquires a
finite polarization correction even at weak coupling, yielding a reduced pairing gap:
\begin{equation}
\Delta^0 (k_F) = \frac{1}{(4e)^{1/3}} \frac{8}{e^2} \frac{\hbar^2 k_F^2}{2m} \exp\left( \frac{\pi}{2ak_F}\right)~.
\label{eq:weakGMB}
\end{equation}
The polarization corrections reduce the mean-field BCS result
by a factor of  $1 / (4e)^{1/3} \approx 0.45$. Interestingly, if one treats the polarization effects at the level of sophistication used in the work of Gorkov and Melik-Barkhudarov, this factor changes with $k_F a$ \cite{Schulze:2001}, though there is no {\it a priori} reason to expect such an approach to be valid at stronger coupling ($k_F a$ of order 1 or more). Calculating the pairing gap in this region has been an onerous task, as can be seen from the multitude of 
publications devoted to this subject in the past few decades.\cite{Lombardo:2001,Dean:2003,Chen:1993,Wambach:1993,Schulze:1996,Schwenk:2003,Muether:2005,Fabrocini:2005,Cao:2006,Margueron:2008,Gandolfi:2008,Abe:2009,Gandolfi:2009}

\subsection{BCS in the continuum and in a box}
\label{sec:bcs}
As the coupling strength increases, we expect the BCS mean-field
theory to become more accurate. In the BCS-BEC transition studied in 
cold atoms, the BCS theory goes correctly to the two-body bound state
equation in the deep BEC regime.
Though we do not expect BCS results to be precise,
BCS theory provides a standard
basis of comparison for our {\it ab initio} results
and also allows us to analyze finite-size effects in the
QMC simulations in a simple way.
Within the BCS formalism the wave function is:
\begin{equation}
| \psi \rangle = \prod_{\bf k} (u_{{\bf k}} + v_{{\bf k}} c^{\dagger}_{{\bf k} \uparrow}
c^{\dagger}_{{-\bf k} \downarrow})|0\rangle~,
\label{eq:bcswave}
\end{equation}
where $u_{{\bf k}}^2 + v_{{\bf k}}^2 = 1$. A variational minimization of the expectation value of the Hamiltonian 
for an average partice number (or density) leads
to the gap equation:
\begin{equation}
\Delta({\bf k}) = -\sum_{{\bf k'}} \langle {\bf k} |v | {\bf k'} \rangle \frac{\Delta({\bf k'})}{2E({\bf k'})}~,
\label{deltageneraleq}
\end{equation}
where the elementary quasi-particle excitations of the system have energy:
\begin{equation}
E({\bf k}) = \sqrt{\xi({\bf k})^2+\Delta({\bf k})^2}
\label{quasienereq}
\end{equation}
and $\xi({\bf k}) = \epsilon({\bf k})-\mu$, where the chemical potential is $\mu$ and $\epsilon({\bf k}) = \frac{\hbar^2k^2}{2m}$ 
is the single-particle energy of a particle with momentum ${\bf k}$. The chemical potential is found by solving the gap equation together with the equation that provides the average particle number:
\begin{equation}
\langle N \rangle = \sum_{{\bf k}} \left [ 1 - \frac{\xi({\bf k})}{E({\bf k})} \right ]~.
\label{particlegeneraleq}
\end{equation}
When interested in the $^1\mbox{S}_0$ gap for neutron matter, it is customary to perform partial-wave expansions of the potential
and the gap functions, as well as an angle-average approximation. Thus, Eq. (\ref{deltageneraleq}) takes the form:
\begin{equation}
\Delta(k) = -\frac{1}{\pi} \int\limits_0^{\infty} dk' ~k'^2 \frac{v(k,k')}{E(k')} \Delta(k')~,
\label{deltaconteq}
\end{equation}
where the potential matrix element is:
\begin{equation}
v(k,k') = \int\limits_0^{\infty} dr ~r^2 j_{0}(k'r) V(r) j_{0}(kr)~.
\end{equation}
Similarly, Eq. (\ref{particlegeneraleq}) becomes:
\begin{equation}
\rho = \frac{1}{2\pi^2} \int\limits_0^{\infty} dk ~k^2 \left(1 - \frac{\xi(k)}{E(k)}\right)~.
\label{particleconteq}
\end{equation}
These equations are one dimensional, and thus simple to treat numerically. The density equation can be decoupled from the gap 
equation only when $\Delta / \mu<<1$; this is not the case for the density regime we are considering. 

\begin{figure}[t]
\vspace{0.5cm}
\begin{center}
\includegraphics[width=0.46\textwidth]{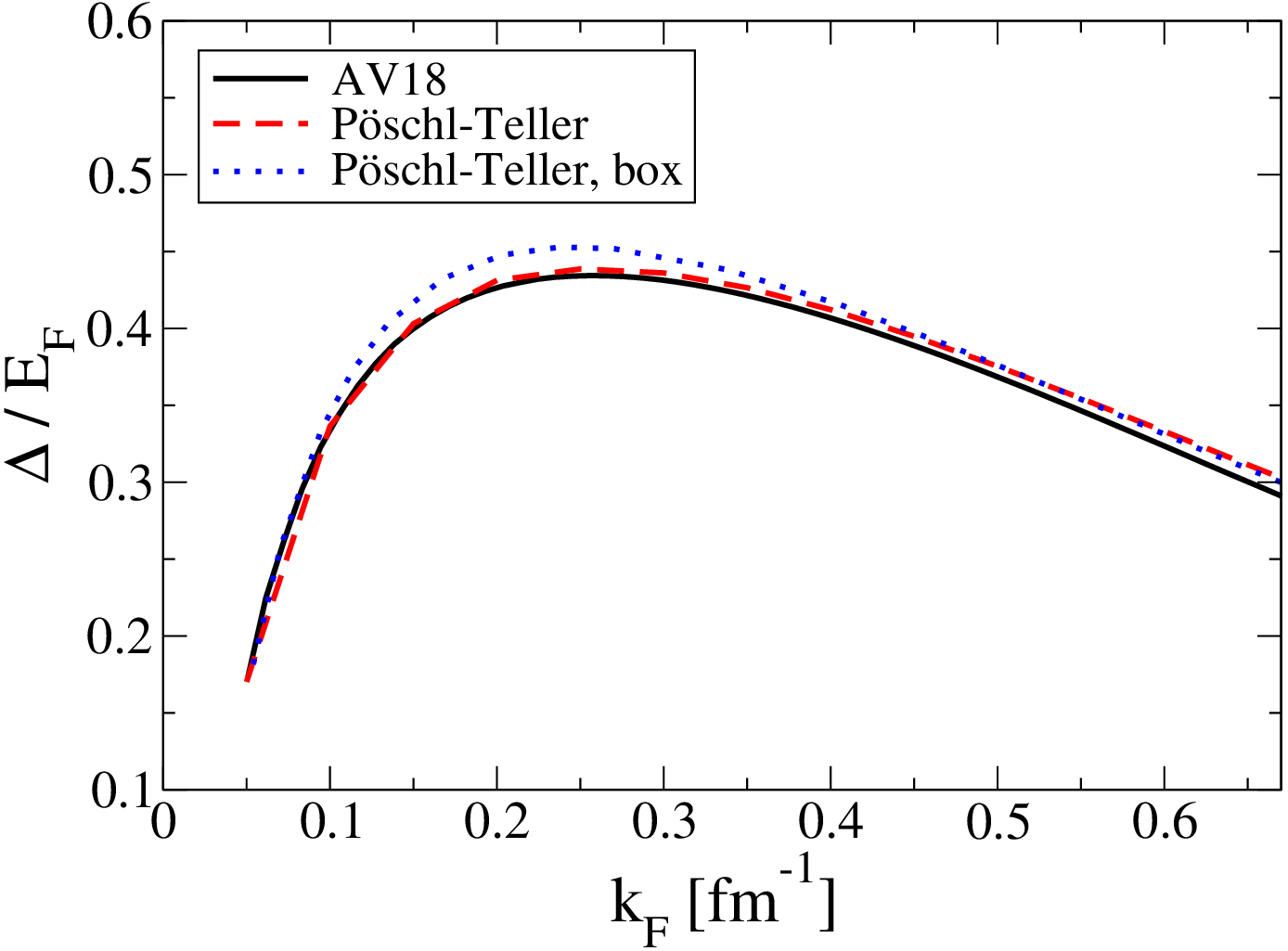}
\caption{(color online) BCS neutron-matter pairing gap $\Delta$ divided with the Fermi energy $E_F$, versus the Fermi momentum $k_F$ for AV18 (solid line) and a modified P\"{o}schl-Teller potential (dashed line) tuned to have the
same scattering length and effective range as AV18. At low density the two curves are identical for practical purposes. Also shown is the solution of the BCS problem in a periodic box using the modified P\"{o}schl-Teller potential for 66 particles (dotted line).}
\label{fig:bcs}
\end{center}
\end{figure}

We have solved Eq. (\ref{deltaconteq}) in tandem with Eq. (\ref{particleconteq}) for the $^1\mbox{S}_0$ channel of the Argonne v18 \cite{Wiringa:1995} potential
that contains a strong short-range repulsion. This calculation is greatly simplified if one uses the method described in Ref. \cite{Khodel:1996}, 
thereby transforming the problem into a quasilinear one. We have also solved Eq. (\ref{deltaconteq}) together with Eq. (\ref{particleconteq}) for 
a modified P\"{o}schl-Teller potential:
\begin{equation}
v(r) = - v_0 \frac{2 \hbar^2}{m} \frac{\nu^2}{\cosh^2(\nu r)}~,
\label{cosheq}
\end{equation}
where $v_0$ and $\nu$ are parameters which we tuned so that this potential reproduces the neutron-neutron scattering length $a \approx -18.5$ fm and effective range $r_e \approx 2.7$ fm.
The potential in Eq. (\ref{cosheq}) clearly has no repulsive core, making it 
more amenable to a straightforward iterative solution. In Fig. \ref{fig:bcs} we show the results for these two potentials. For all the 
densities considered in this work, the results of solving Eqs. (\ref{deltaconteq}) and (\ref{particleconteq}) with these two 
potentials are virtually indistinguishable.  For treatments beyond the  mean field, though, more care must be taken. A simple purely attractive
interaction with finite positive effective range will produce a collapse to a system size of the range of the potential.  The repulsive core avoids
this collapse in QMC calculations, but the details of the core interaction are not important at the low densities considered here. 

The modified P\"{o}schl-Teller potential can also be used in a calculation for finite average particle number. We have solved Eqs. (\ref{deltageneraleq}) and (\ref{particlegeneraleq}) for $\langle N \rangle$ from 20 to 200, 
in periodic boundary conditions in a cubic box of volume $L^3$:
\begin{equation}
{\bf k}_{\bf n} = \frac{2 \pi}{L} ( n_{x}, n_{y}, n_{z})~.
\label{borneq}
\end{equation}
The solution to this problem for many values of $\langle N \rangle$ at a fixed density of $k_F a = -10$
was given in Ref. \cite{Gezerlis:2008}. There it was found that $\langle N \rangle = 66$ is very close to the thermodynamic limit. 
We have performed similar calculations for other densities, finding that they all exhibit the same trend. We have also performed similar computations with generalized boundary conditions including separate phase shifts for
spin up and down particles at the box boundary.\cite{Gezerlis:2009}
In Fig. \ref{fig:bcs} we show the results of solving the BCS gap equation Eq. (\ref{deltageneraleq}) in a periodic box 
along with the particle-number conserving Eq. (\ref{particlegeneraleq}) for $\langle N \rangle = 66$. 
We do not expect this procedure to be sufficient for very weak coupling, $| k_F a | < 1$, as  the pair size becomes larger than the simulation volume.  A more detailed study is warranted to attempt to extract pairing gaps in this regime. In the rest of this work, when we mention BCS results in a box 
we will refer to the $\langle N \rangle = 66$ case.

We have also calculated, both in the continuum and in a periodic box, the momentum distribution, which in BCS is given by the following expression
\begin{equation}
 n({\bf k}) = \frac{1}{2} \left [ 1 - \frac{\xi({\bf k})}{E({\bf k})} \right ]~,
\end{equation}
as well as the energy of the quasi-particle excitations, which follows from Eq. (\ref{quasienereq}). These are given in sections \ref{sec:gap} and \ref{sec:distrib}.

\section{Quantum Monte Carlo}
\label{sec:qmc}
\subsection{Hamiltonian}
\label{sec:hamilt}

The Hamiltonian for neutron matter at low densities is:
\begin{equation}
{\cal{H}} = \sum\limits_{k = 1}^{N}  ( - \frac{\hbar^2}{2m}\nabla_k^{2} )  + \sum\limits_{i<j'} v(r_{ij'})~.
\end{equation}
where $N$ is the total number of particles. The neutron-neutron interaction is generally quite complicated, having one-pion exchange at large distances, an intermediate 
range spin-dependent attraction by two-pion exchange, and a short-range repulsion. In the regime of interest, though, the scattering 
length and the effective range are most crucial to the physical properties of the system. For the purposes of our simulation, a short-range repulsive 
core is also important so as to avoid a collapse to a higher-density state. 

In Ref. \cite{Gezerlis:2008} we used the $^{1}\mbox{S}_0$ potential as the interaction between all opposite-spin pairs. A perturbative correction 
was added to correct for the fact that the $S=1, M_S=0$ pairs must be in a relative $p$-state (or higher) due to anti-symmetry.  The $p$-wave interaction was
neglected in those calculations.
Here we improve this treatment by explicitly including $p$-wave interactions in the same-spin pairs, and perturbatively correcting the $S=1, M_S=0$
pairs to the correct $p$-wave interaction. We use the AV4 potential to determine the $p$-wave interactions.\cite{Wiringa:2002}.

The AV4 interaction for neutrons can be simplified to
\begin{equation}
v_4(r) = v_c(r) + v_{\sigma}(r){\mbox{\boldmath$\sigma$}}_1\cdot{\mbox{\boldmath$\sigma$}}_2,
\label{vfour}
\end{equation}
which in the case of the S=0 singlet pairs gives 
\begin{equation}
v_S(r) = v_c(r) - 3v_{\sigma}(r)~.
\label{ves}
\end{equation}
In this paper we add the contribution from spin 1 (triplet) pairs:
\begin{equation}
v_P(r) = v_c(r) + v_{\sigma}(r)~.
\end{equation}
The same-spin potential contribution is small even at the highest density considered.
While still keeping the potential of Eq. (\ref{ves}) in the propagator of our QMC method for the opposite-spin pairs,
we have corrected perturbatively using the general case described in Eq. (\ref{vfour}), 
which can be written as (see, e.g., Ref. \cite{Blatt:1952}):
\begin{equation}
v_4(r) = v_c(r) + v_{\sigma}(r)(-2P^M - 1)
\label{vmajorana}
\end{equation}
in terms of the Majorana exchange operator.

\subsection{Variational and Green's Function Monte Carlo}
We employ standard Variational and Green's function Monte Carlo methods to calculate the properties
of dilute neutron matter.  VMC calculations use Monte Carlo integration
to minimize the expectation value of the Hamiltonian:
\begin{equation}
\langle H \rangle_{VMC} = \frac{\int d{\bf R} \Psi_{V}({\bf R}) H \Psi_{V}({\bf R})}{\int d{\bf R} |\Psi_{V}({\bf R})|^2} \geq E_0~.
\label{eq:qmcvmc}
\end{equation}
thereby optimizing the variational wave function $\Psi_V$.

Fixed-node GFMC simulations  project out the lowest-energy eigenstate $\Psi_{0}$ from a trial (variational) wave function $\Psi_{V}$. This they do by treating the Schr\"{o}dinger equation as a diffusion equation in imaginary time $\tau$ and evolving the variational wave function up to large $\tau$.

The ground state is evaluated from:
\begin{eqnarray}
\Psi_0 & = & \exp [ - ( H - E_T ) \tau ] \Psi_V  \\ \nonumber
& = & \prod \exp [ - ( H - E_T ) \Delta \tau ] \Psi_V,
\end{eqnarray}
evaluated as a branching random walk.  The short-time propagator
is usually taken as
\begin{eqnarray}
\exp [ -H \Delta \tau ] = \exp [ -V \Delta \tau / 2 ] \exp [ -T \Delta \tau ] \exp [ -V \delta \tau / 2 ],
\label{eq:shorttime}
\end{eqnarray}
which is accurate to order $(\Delta \tau)^2$.
For the lowest densities considered, we include the two-body 
propagator exactly:
\begin{equation}
\exp [ -H \Delta \tau ]  = \exp [ - T \Delta \tau ] 
\frac { \exp [ -H_2 \Delta \tau ] }{\exp [ -H_0 \Delta \tau ]},
\end{equation}
where the two-body Hamiltonian $H_2 $  includes the pair relative kinetic
energy and the pair potential and $H_0$ is the pair kinetic term only.
At lowest order in $(\Delta \tau)$ this is equivalent to Eq. \ref{eq:shorttime}.
However it includes multiple scattering for a pair and allows accurate 
calculations with larger time steps $\Delta \tau$.  This is particularly important
for very dilute systems where these multiple-scattering contributions of
individual pairs dominate.

The fixed node calculation gives
a wave function $\Psi_0$ that is the lowest-energy state with the sames
nodes (surface where $\Psi$ = 0) as the trial state $\Psi_V$.  The resulting
energy $E_0$ is an upper bound to the true ground-state energy.
The variational wave function $\Psi_V$  has a Jastrow-BCS form
(see next sub-section), and contains a variety of parameters, many of
which affect the nodal surfaces. Since the
fixed-node energy is an upper bound to the true ground state, these
parameters can be optimized to give the best approximation 
to the ground-state wave function. In order to optimize
these variational parameters, we include them as slowly diffusing
coordinates in a preliminary GFMC calculation.  The parameters
evolve slowly in imaginary time, equilibrating around the lowest-energy
state consistent with the chosen form of the trial wave function.\cite{Carlson:2003}

The ground-state energy $E_0$ can be obtained from:
\begin{equation}
E_0 = \frac{ \langle \Psi_V | H | \Psi_0 \rangle}{ \langle \Psi_V | \Psi_0 \rangle}
= \frac{ \langle \Psi_0 | H | \Psi_0 \rangle}{ \langle \Psi_0 | \Psi_0 \rangle}.
\end{equation}
Expectation values of quantities that do not commute with the Hamiltonian can be calculated using a combination of the mixed and variational estimate:
\begin{equation}
\langle \Psi_0 | \hat{S} | \Psi_0 \rangle \approx 2 \langle \Psi_0 | \hat{S} | \Psi_V \rangle - \langle \Psi_V | \hat{S} | \Psi_V \rangle~,
\label{eq:extrapo}
\end{equation}
where $\hat{S}$ is the operator corresponding to the relevant physical quantity, and the error in this expression is of second order in $\Psi_0 - \Psi_V$. Such a combination of estimates is often called the ``extrapolated estimate''.

\subsection{Trial wave function}
\label{sec:wave}

\begin{figure}[t]
\vspace{0.5cm}
\begin{center}
\includegraphics[width=0.46\textwidth]{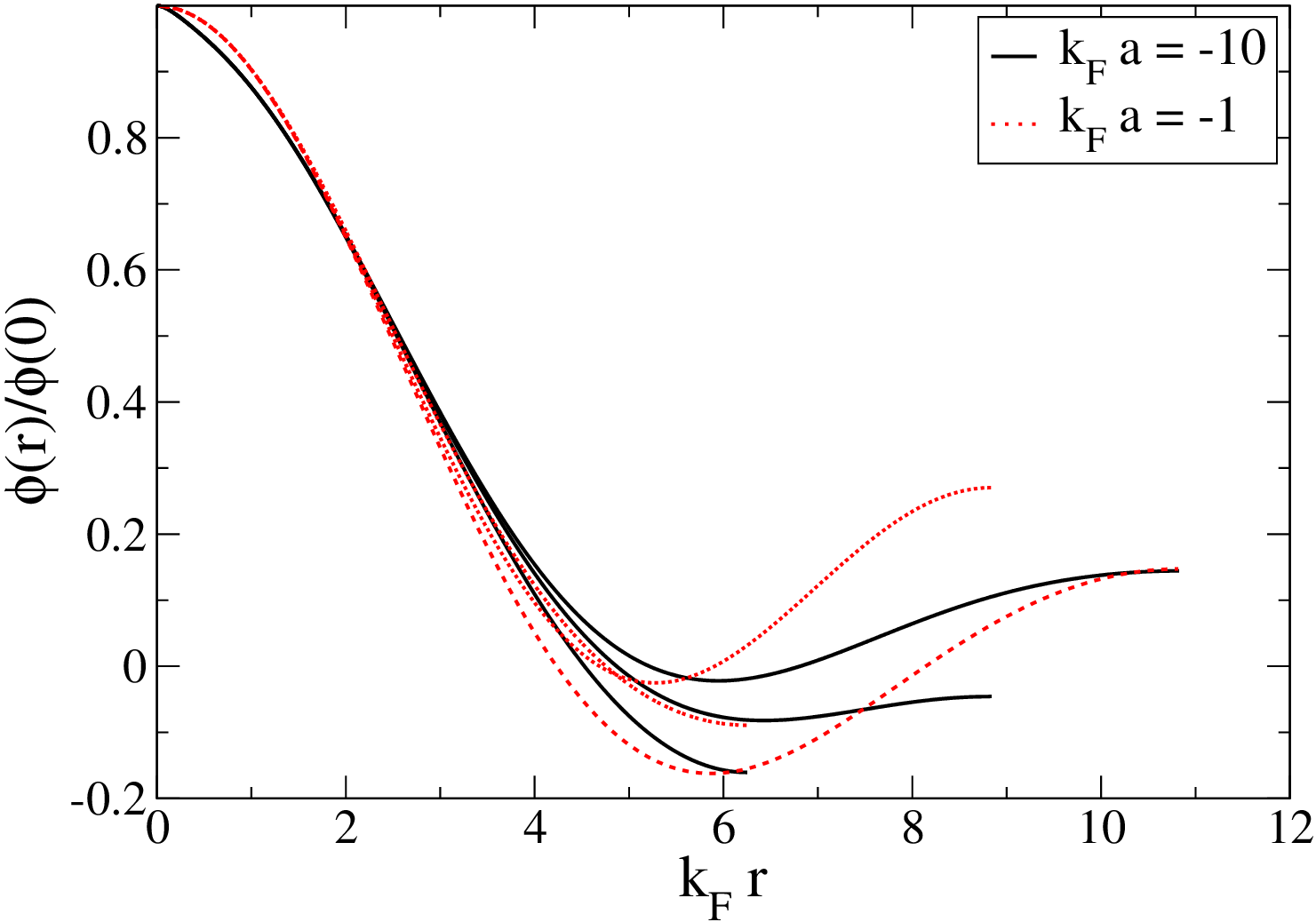}
\caption{(color online) Normalized neutron-matter variationally optimized pairing function $\phi(r)$ for $k_Fa = -10$ (solid lines) and $k_Fa = -1$ (dotted lines), for different directions in the periodic simulation volume (in terms of rising expanse they correspond to the 001, 011, and 111 directions in the box).}
\label{fig:phi}
\end{center}
\end{figure}

We take the trial wave function to be of the Jastrow-BCS form with fixed particle number:
\begin{equation}
\Psi_V = \prod\limits_{i \neq j} f_P(r_{ij}) \prod\limits_{i' \neq j'} f_P(r_{i'j'}) \prod\limits_{i,j'} f(r_{ij'})  {\cal A} [ \prod_{i<j'} \phi (r_{ij'}) ]
\end{equation}
and periodic boundary conditions. The primed (unprimed) indices correspond to spin-up (spin-down) neutrons. The pairing function
$\phi (r)$ is a sum over the momenta compatible with the periodic boundary conditions. In the BCS theory the pairing function is:
\begin{equation}
\phi(r) =\sum\limits_{\bf n} \frac{v_{{\bf k}_{\bf n}}}{u_{{\bf k}_{\bf n}}} e^{i{\bf k}_{\bf n}\cdot{\bf r}} =\sum_{\bf n} \alpha_n e^{i{\bf k}_{\bf n}\cdot{\bf r}} ~,
\end{equation}
and here it is parametrized with a short- and long-range part as in Ref. \cite{Carlson:2003}:
\begin{equation}
\phi({\bf r}) = \tilde{\beta} (r) + \sum_{{\bf n},~I \leq I_C} \alpha_I e^{ i {\bf{k}}_{\bf n} \cdot {\bf r}}~,
\label{phieq}
\end{equation}
where $I = n_x^2 + n_y^2 + n_z^2$ using the parameters defined in Eq. (\ref{borneq}). The Jastrow part is taken from a lowest-order-constrained-variational method \cite{Pandharipande:1973} calculation described by a Schr\"{o}dinger-like equation:
\begin{equation}
- \frac{\hbar^2}{m}\nabla^{2} f(r)  + v(r) f(r) = \lambda f(r)~\nonumber
\end{equation}
for the opposite-spin $f(r)$ and by the corresponding equation for the same-spin $f_P(r)$. Since the $f(r)$ and $f_P(r)$ we get are nodeless, they do not affect the final result apart from reducing the statistical error. The fixed-node approximation
guarantees that the result we obtain for one set of pairing function parameters in Eq. (\ref{phieq}) will be an upper bound to the true ground-state energy
of the system. As in previous works \cite{Carlson:2003, Gezerlis:2008} the parameters are optimized in the full QMC calculation, providing the
best possible nodal surface, in the sense of lowest fixed-node energy, with the given form of trial function.
We utilize this upper-bound property to get as close to the true ground-state energy as possible.

Given the finite-size analysis shown in Refs. \cite{Gezerlis:2008,Gezerlis:2009}, we have performed all calculations for 66-68 particles in periodic
boundary conditions. The equation of state is determined from the 66 particle results, and the pairing gap from the odd-even staggering.
We have separately optimized the wave-function parameters at each density, and show the results for $\phi (r)$ (normalized each time to the value at zero separation) for the largest and smallest density we have considered ($k_Fa = -10$ and $k_Fa = -1$) in Fig. \ref{fig:phi}.

\subsection{Equation of state}
\label{sec:eos}

\begin{figure}[t]
\vspace{0.5cm}
\begin{center}
\includegraphics[width=0.46\textwidth]{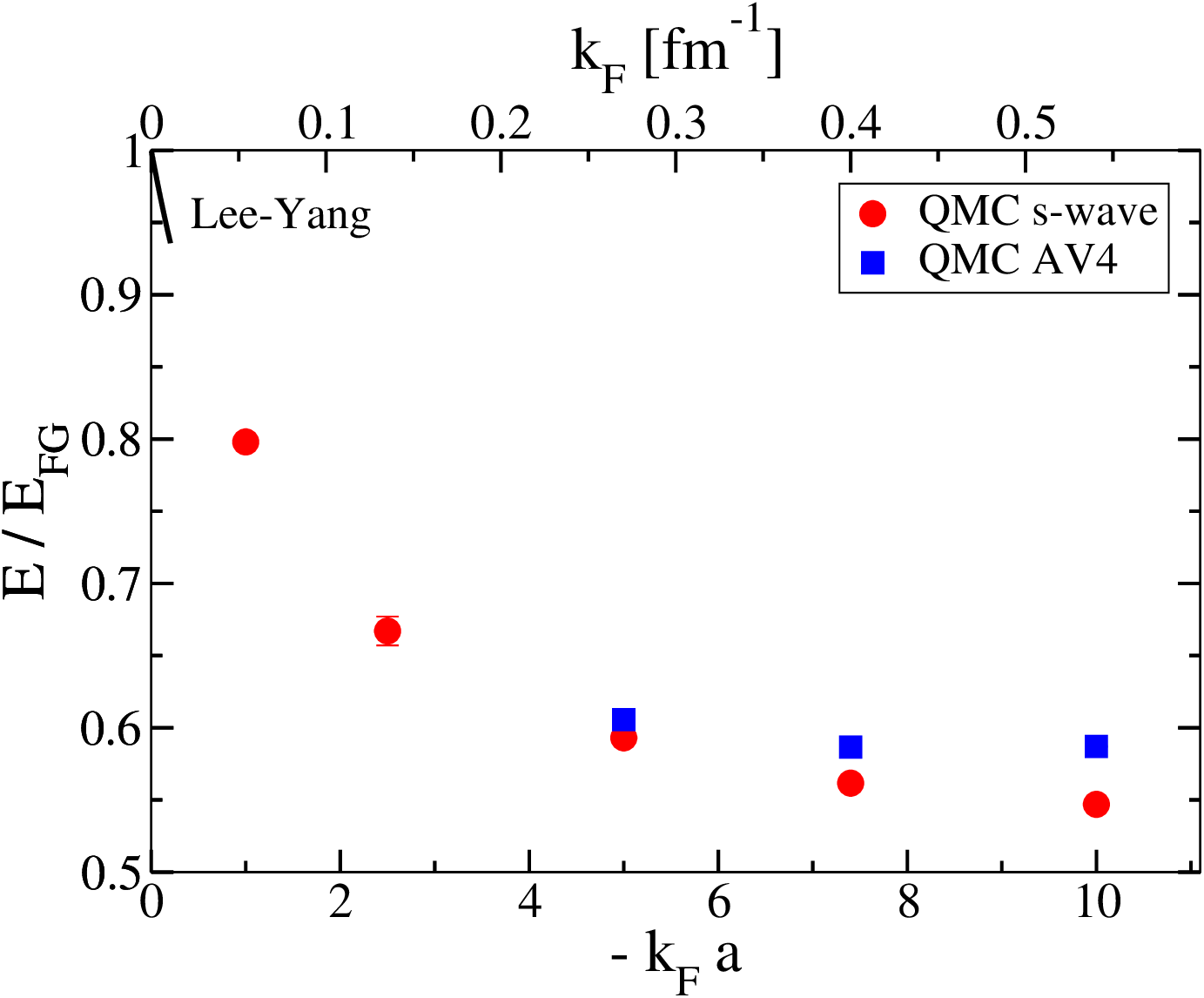}
\caption{(color online) Equation of state for neutron matter using different potentials. Shown are QMC results for the $s$-wave potential (circles) and for the AV4 (squares). Also shown is the analytic expansion of the ground-state energy of a normal fluid (line).}
\label{fig:AFDMClikeener}
\end{center}
\end{figure}

We first examine the equation of state of low-density neutron matter,  in particular its evolution from the weak- to strong-coupling regime
and the impact of adding $p$-wave interactions between the neutrons.
In Fig. \ref{fig:AFDMClikeener} we show the equation of state for low-density neutron matter with the  $s$-wave interaction potential along with the new AV4 results. It is clear that when the density is very low, the $s$-wave contribution is dominant, and our results for the lowest densities remain unchanged. At higher densities the energy is higher with the contribution of the $p$-wave interaction. For the highest density examined, $k_F a = -10$, this change is approximately 7\%, while for $k_F a = -5$ it is  only 1\%. Nonperturbative corrections at the highest density considered could reduce the difference between the $s$-wave interaction and AV4 results slightly. The curve at lower densities shows the analytic result \cite{Lee:1957} described in section \ref{sec:exact}. Our calculations extend to lower densities than other microscopic calculations, and agree with the trend implied by the Lee-Yang result.

\begin{figure}
\vspace{0.5cm}
\begin{center}
\includegraphics[width=0.46\textwidth]{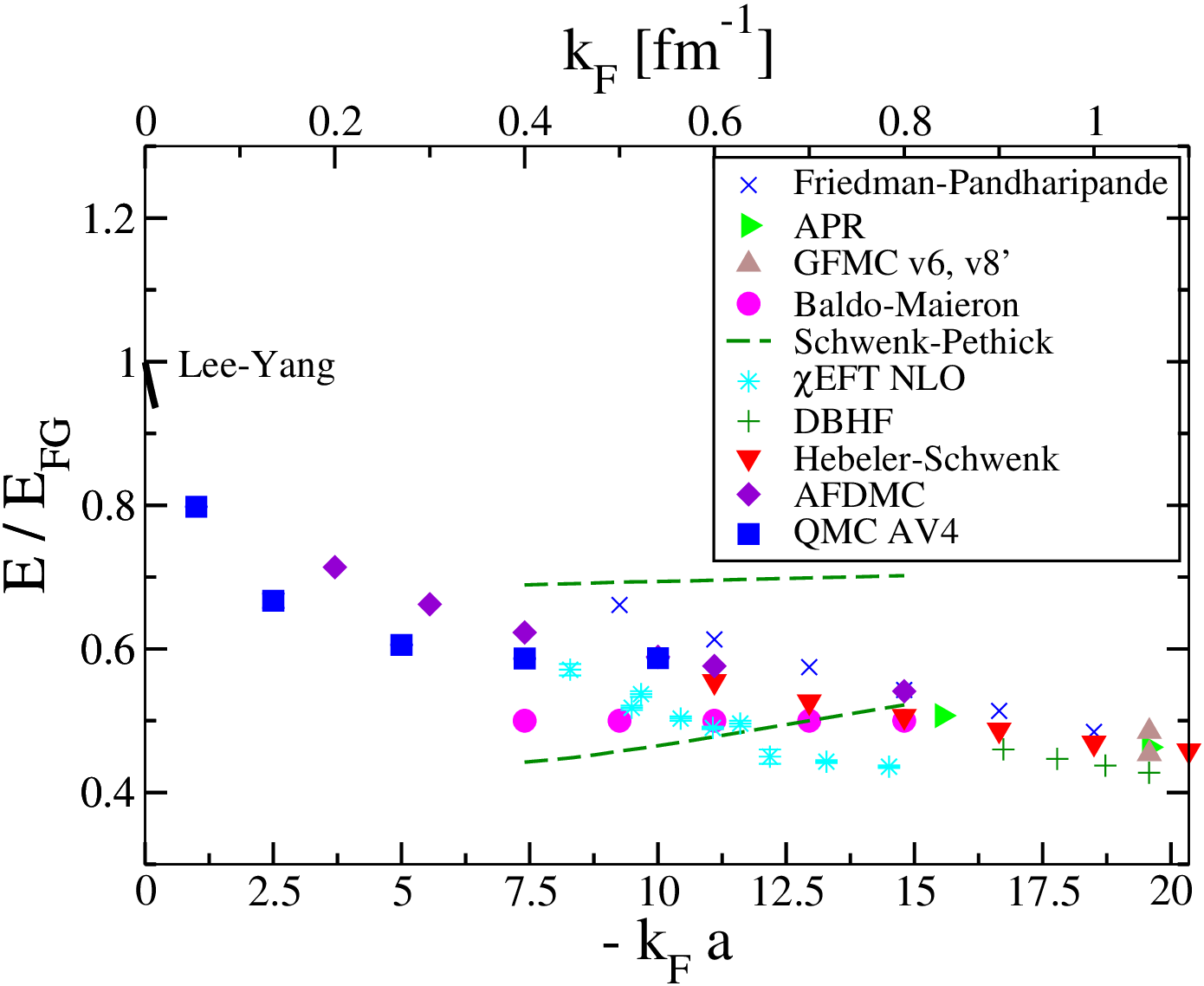}
\caption{(color online) Equation of state for neutron matter compared to various previous results. Despite quantitative discrepancies, all calculations give essentially similar results. Our lowest density corresponds to $k_F a = -1$.}
\label{fig:DL}
\end{center}
\end{figure}

We also compare our QMC AV4 results from Fig. \ref{fig:AFDMClikeener} for the ground-state energy to other calculations extending to larger Fermi momenta. In Fig. \ref{fig:DL} we compare our results to (approximate) variational hypernetted-chain calculations by Friedman and Pandharipande \cite{Friedman:1981}, and another calculation by Akmal, Pandharipande, and Ravenhall (APR) \cite{Akmal:1998}. We also include two Green's Function Monte Carlo results for 14 neutrons with more complete Hamiltonians \cite{Carlson:Morales:2003}, a result following from a Brueckner-Bethe-Goldstone expansion \cite{Baldo:2008}, a difermion EFT result (shown are the error bands) \cite{Schwenk:2005}, the latest Auxiliary-Field Diffusion Monte Carlo (AFDMC) result (discussed below) \cite{Gandolfi:2008}, a Dirac-Brueckner-Hartree-Fock calculation \cite{Margueron:2007b}, a lattice chiral EFT method at next to leading order \cite{Epelbaum:2008b} (see also Ref. \cite{Epelbaum:2008a}), and an approach that makes use of chiral N$^2$LO three-nucleon forces.\cite{Hebeler:2009} Of these, Refs. \cite{Akmal:1998}, \cite{Gandolfi:2008}, and \cite{Hebeler:2009} include a three-nucleon interaction, though at the densities we consider, these are not expected to be significant.
Qualitatively all of these results agree within 20\%.  

A series of {\it ab initio} calculations for neutron matter using the AFDMC method have been
published beginning in 2005.\cite{Fabrocini:2005} After our analysis of the finite-size effects -- described for BCS in section \ref{sec:bcs} and for QMC in Refs. \cite{Gezerlis:2008,Gezerlis:2009} -- was published in late 2007, the AFDMC group repeated their calculations for larger systems, \cite{Gandolfi:2008,Gandolfi:2009} bringing them closer to our results, though still, as can be seen from Fig. \ref{fig:DL} the results are distinct. Given the {\it ab initio} nature of the powerful AFDMC method, \cite{Schmidt:1999} we have attempted to compare results more extensively. The advantage of the AFDMC approach is that it
includes an interaction which is more complete than the simpler ones used here. The disadvantage
of the AFDMC approach is that it does not provide a variational bound
to the energy, and hence the wave functions are chosen from another approach.
In the calculations of Refs. \cite{Fabrocini:2005,Gandolfi:2008,Gandolfi:2009} the wave function was taken from a Correlated-Basis
Function (CBF) approach that included a BCS-like initial state. The pairing
in that variational state is unusually large, and in fact increases as a fraction
of $E_F$ when the density is lowered. 

The QMC AV4 results use a wave function that has been variationally optimized. QMC thus gives energies that are considerably lower than the AFDMC results. As both the wave functions and the interactions are different in the previous QMC and AFDMC results,
we have repeated our calculations using the same input wave function \cite{Gandolfi:pc} used by the AFDMC group (which comes from the same Correlated-Basis Function calculation) at $k_F = 0.4 ~\mbox{fm}^{-1}$ and at $k_F a = -10$. We find that in QMC the AV4 results for the optimized wave function [0.5866(6) MeV and 0.5870(3) MeV, respectively] are consistently lower in energy than those using the CBF as input [0.6254(9) MeV and 0.6014(7) MeV, respectively].
This means that they are closer to the true ground-state energy for the Hamiltonian we consider. It would be worth studying in more detail the differences arising from the different Hamiltonians; the most important remaining differences are likely the spin-orbit and pion-exchange terms in the $p$-wave interaction. Extensions of previous GFMC calculations \cite{Carlson:Morales:2003} to lower densities would help to resolve these issues.
 
It is interesting to note that at the lowest densities considered, the AFDMC and QMC results are still distinct. At those densities contributions of $p$- and higher partial waves  in the Hamiltonian should be very small, and thus the two methods should give identical results. The three-nucleon interaction included in the AFDMC calculations  is one possible source of the difference, though this appears unlikely at the smallest densities considered.  This suggests that the CBF wave function at very low densities is problematic; additional studies with Jastrow-BCS or other wave functions would be useful.

\subsection{Pairing gap and quasiparticle spectrum}
\label{sec:gap}

\begin{figure}[b]
\vspace{0.5cm}
\begin{center}
\includegraphics[width=0.46\textwidth]{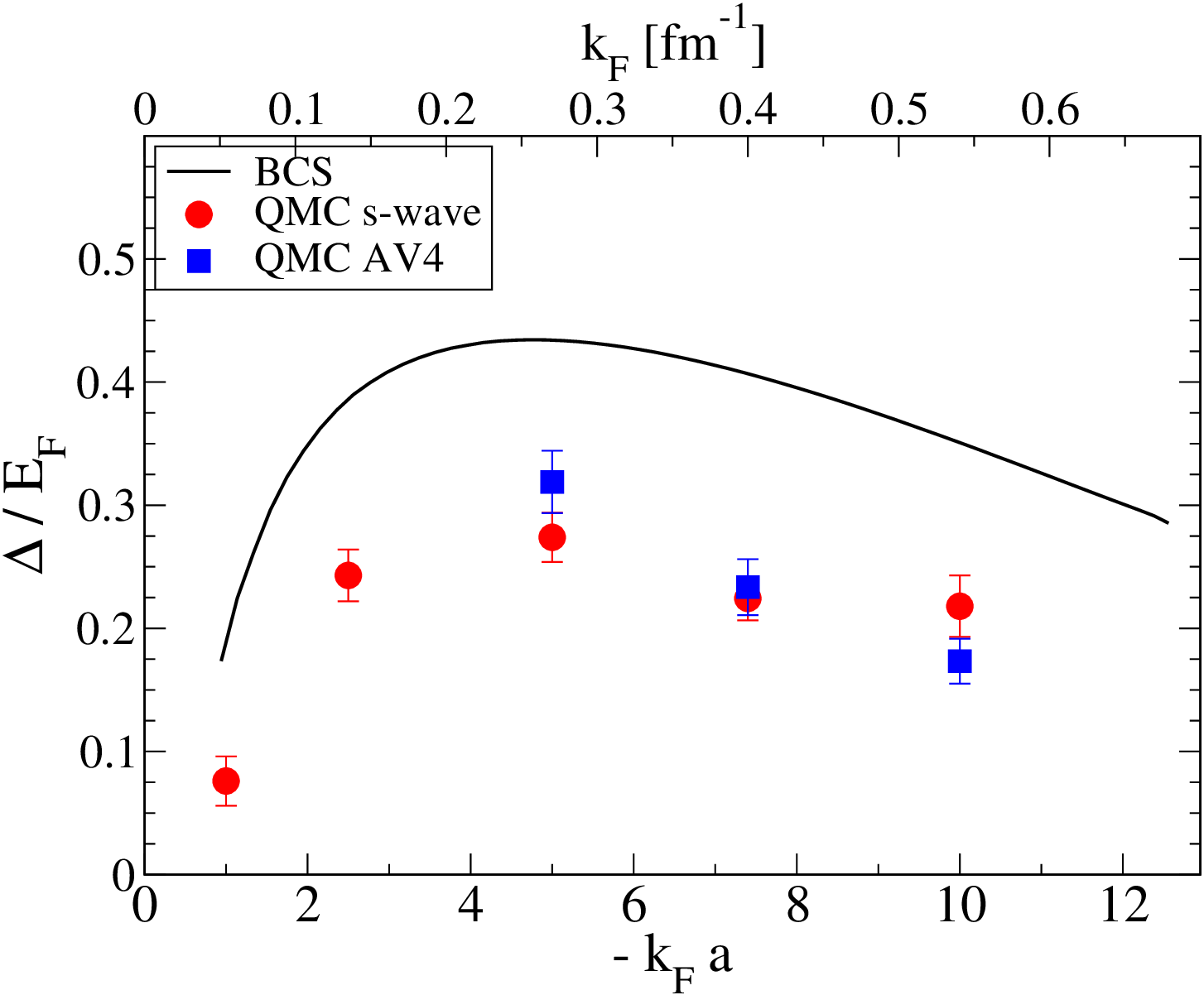}
\caption{(color online) Superfluid pairing gap versus $k_F a$ for neutron matter using different potentials. Shown are QMC results for the $s$-wave potential (circles) and for the AV4 (squares). Also shown is the mean-field BCS result (line).}
\label{fig:AFDMClikegap}
\end{center}
\end{figure}

We have also performed calculations for the zero-temperature pairing gap using the AV4 interaction. These follow from our knowledge of the ground-state energy, through the use of the the odd-even staggering formula:
\begin{equation}
\Delta = E(N+1) - \frac{1}{2} \left [ E(N)+E(N+2) \right ]~,
\label{eq:staggerer}
\end{equation}
where $N$ is an even number of particles. The results for the gap are shown in Fig. \ref{fig:AFDMClikegap}. The main conclusion is that the gap remains essentially unchanged with the inclusion of the $p$-wave interactions. Even at the highest density examined, $k_F a = -10$, the gap is within statistical errors the same comparing $s$-wave and AV4 interactions. This implies that the dominant contributions to the gap come from the $s$-wave part of the interaction.

Our results indicate that the gap is suppressed by approximately a factor of two from the BCS value at $k_F a = -1$, roughly consistent with the Gorkov and Melik-Barkhudarov, Eq. (\ref{eq:weakGMB}), polarization suppression. In cold atoms, this suppression from BCS is reduced as the density increases, with a smoothly growing fraction
of the BCS results as we move from the BCS to the BEC regime.  At unitarity the measured pairing gaps \cite{Shin:2008,Carlson:2008,Schirotzek:2008} are 0.45(0.05)
of the Fermi energy, for a ratio $\Delta/\Delta_{BCS} \approx 0.65$,  in agreement with predictions by QMC methods.\cite{Carlson:2003,Carlson:2005,Gezerlis:2008} In neutron matter, though, the
finite range of the potential reduces $\Delta/E_F$ as the density increases.  We find a ratio $\Delta/\Delta_{BCS}$ that increases slightly from $|k_F a| = 1$ to $2.5$, but thereafter remains roughly constant.

\begin{figure}[b]
\vspace{0.5cm}
\begin{center}
\includegraphics[width=0.46\textwidth]{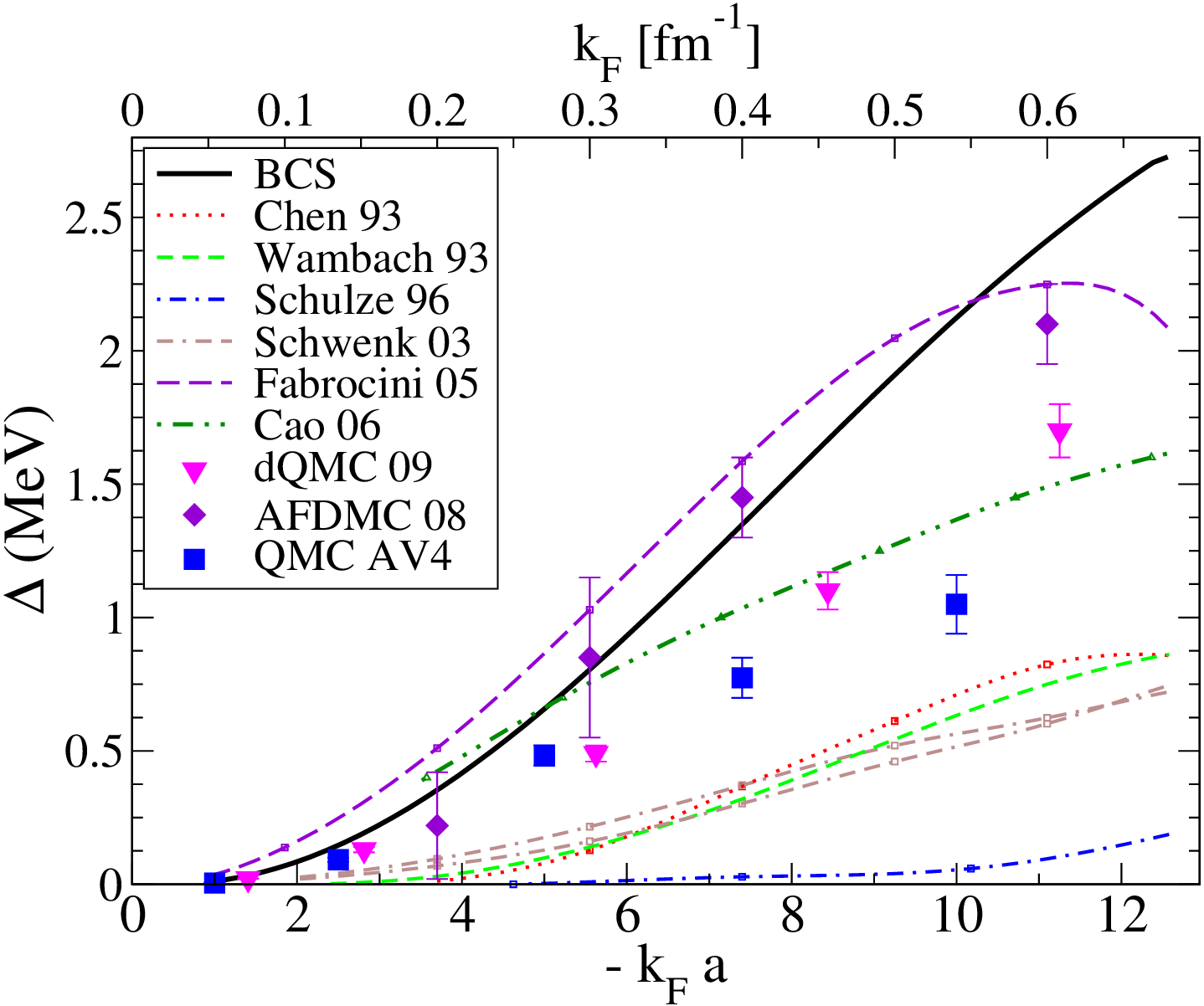}
\caption{(color online) Superfluid pairing gap versus $k_F a$ for neutron matter compared to previous results.}
\label{fig:gap-compare}
\end{center}
\end{figure}

We also used our AV4 calculations to compute the difference between $s$-wave and AV4 interaction
gaps in perturbation theory, in an attempt to isolate
the effects of the addition of the $p$-wave interaction. This perturbation
theory may not be accurate for the highest density considered, since the $s$-wave and
AV4 ground states are somewhat different in energy. It should give
an accurate picture at lower densities, though, and in particular isolate
the sign of the change arising from the $p$-wave terms in the interaction.
Using perturbation theory yields much smaller statistical errors than
comparing the separate $s$-wave and AV4 calculations.
Table I shows that the $p$-wave interactions increase the
pairing gap modestly over the range of densities considered. The $p$-wave
interactions apparently decrease the magnitude of the polarization corrections,
though the change is only approximately 15 \% at the highest density 
considered.

\begin{table}[t]
\caption{Gap differences at various $k_F a$ calculated in perturbation
theory. Perturbative estimates based on AV4 calculations.}
\begin{tabular}{c  c  c  c}
\hline
$k_F a$ & $k_F$ [fm$^{-1}$] &  $ \Delta$(AV4) [MeV] & $\Delta$(AV4) - $\Delta(s)$ [MeV]   \\
\hline
-5.0 & 0.27 & 0.48 (0.04) & 0.012 (0.008) \\
-7.5 & 0.40  & 0.77 (0.08) & 0.11 (0.03) \\
-10.0 & 0.54 & 1.05 (0.11) & 0.16 (0.06) \\
\hline
\end{tabular}
\label{tab:pertgap}
\end{table}

In Fig. \ref{fig:gap-compare} we compare our results to selected previous results: a Correlated-Basis Function calculation by Chen {\it et al.} \cite{Chen:1993}, an extension of the polarization-potential model by Wambach {\it et al.} \cite{Wambach:1993}, a medium-polarization calculation by Schulze {\it et al.} \cite{Schulze:1996}, a renormalization group calculation by Schwenk {\it et al.} \cite{Schwenk:2003}, a Brueckner calculation by Cao {\it et al.} \cite{Cao:2006}, a determinantal lattice QMC approach \cite{Abe:2009}, and finally the newer Correlated-Basis Function calculation by Fabrocini {\it et al.} \cite{Fabrocini:2005} that was used as an input wave function in the two AFDMC calculations of 2005 and 2008.\cite{Fabrocini:2005,Gandolfi:2008} 

The results of our calculations are much larger than most diagrammatic  \cite{Chen:1993,Wambach:1993,Schulze:1996} and renormalization group \cite{Schwenk:2003} approaches. As these approaches assume a well-defined Fermi surface or calculate polarization corrections based on single-particle excitations it is not clear how well they can describe neutron matter in the strongly paired regime, or the similar pairing found in cold atoms. Ref. \cite{Cao:2006}, which incorporates self-energy corrections and screening at the RPA level within Brueckner theory, appears to give results similar to ours. However, these values disagree with our lower-density results and, perhaps more importantly, at the lowest density reported the gap is larger than the mean-field BCS value (see section \ref{sec:exact}). On a similar note, Refs. \cite{Wambach:1993} and \cite{Schwenk:2003} make use of a weak-coupling formula to calculate the pairing gap, similarly to the Eqs. (\ref{eq:weakBCS}) and (\ref{eq:weakGMB}) we discussed in section \ref{sec:exact}. The prefactor they use is justified based on predictions in the theory of $^3$He. However, the concept itself of a Fermi surface is not well-defined in these strongly paired systems: in $^3$He, in contrast to the present case, the gap is considerably smaller than the Fermi energy. 

\begin{figure}[t]
\vspace{0.5cm}
\begin{center}
\includegraphics*[width=0.46\textwidth]{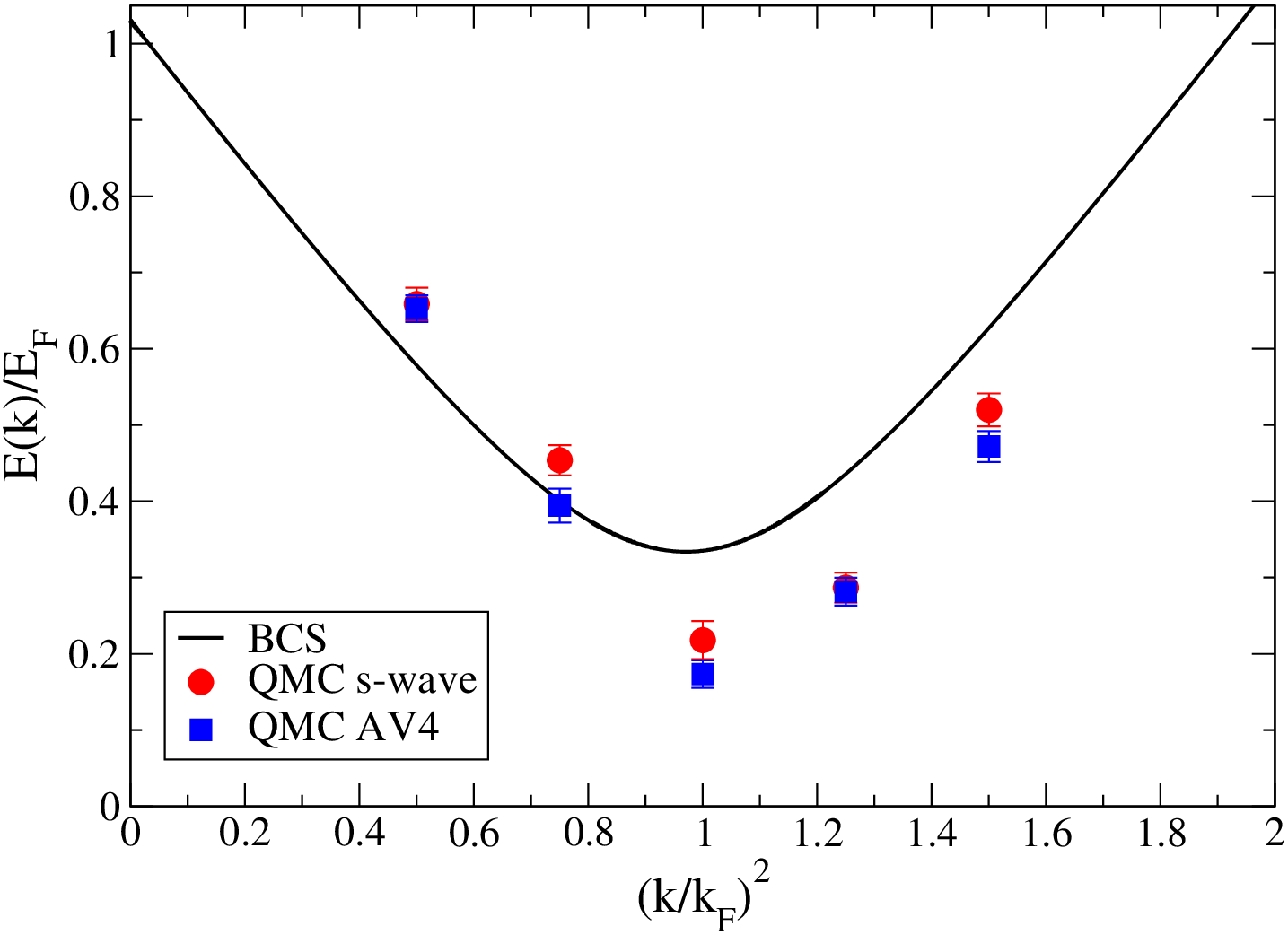}
\caption{(color online) The neutron-matter energy of the quasiparticle excitations of the system in QMC AV4 (squares) versus ${(k/k_F)}^2$ at $k_F a = -10$. Also shown are the BCS continuum results (line) as well as the QMC quasiparticle spectrum that follows from a simple $s$-wave Hamiltonian (circles).}
\label{fig:NMquasi}
\end{center}
\end{figure}

Our results are also somewhat different from the AFDMC results of Ref. \cite{Gandolfi:2008}. We have once again repeated our QMC calculations for the gap using the CBF wave function as input. We find something quite interesting: the QMC method using the AV4 potential and the CBF input wave function at $k_F = 0.4~\mbox{fm}^{-1}$ (which gave an energy higher than the variationally optimized input wave function, see section \ref{sec:eos}) gives a gap of 1.21(17)  MeV, thus reproducing the AFDMC result, which uses the same input wave function and the much more complicated AV8'+UIX interaction, this being 1.45(15) MeV. This too suggests that the most important contributions to the gap come from the $s$-wave part of the interaction. 
On the other hand, our results seem to qualitatively agree (at least for the lowest densities considered) with a determinantal Quantum Monte Carlo lattice calculation \cite{Abe:2009} which, however, only includes the $s$-wave component in the interaction. This may imply that a consensus is emerging, in that both these calculations find a gap that is suppressed with respect to the mean-field BCS result but is still a substantial fraction of the Fermi energy. Finally, let us note that, as mentioned before when discussing Fig. \ref{fig:AFDMClikegap}, the AV4 results for the optimized wave function are very similar to those using an $s$-wave potential. 

We have also calculated the quasiparticle excitation spectrum using the AV4 interaction. The minimum of these results provides the pairing gap in Fig. \ref{fig:AFDMClikegap}. In Fig. \ref{fig:NMquasi} we show both the $s$-wave Hamiltonian results, as well as the AV4 results. In cold atoms at unitarity and
beyond (the BEC regime) the quasiparticle minimum energy is at a momentum significantly smaller than the Fermi momentum.  Here, though, the minimum corresponds closely to the neutron Fermi momentum. Although the QMC minimum (pairing gap) is much smaller than the BCS minimum, the dispersion around the minima is quite similar. Just like in the case of cold atoms, microscopic results for the quasiparticle energy spectra \cite{Carlson:2005} can be used to constrain density functional calculations.\cite{Bulgac:2008}

\subsection{Distribution functions}
\label{sec:distrib}

\begin{figure}[t]
\vspace{0.5cm}
\begin{center}
\includegraphics*[width=0.46\textwidth]{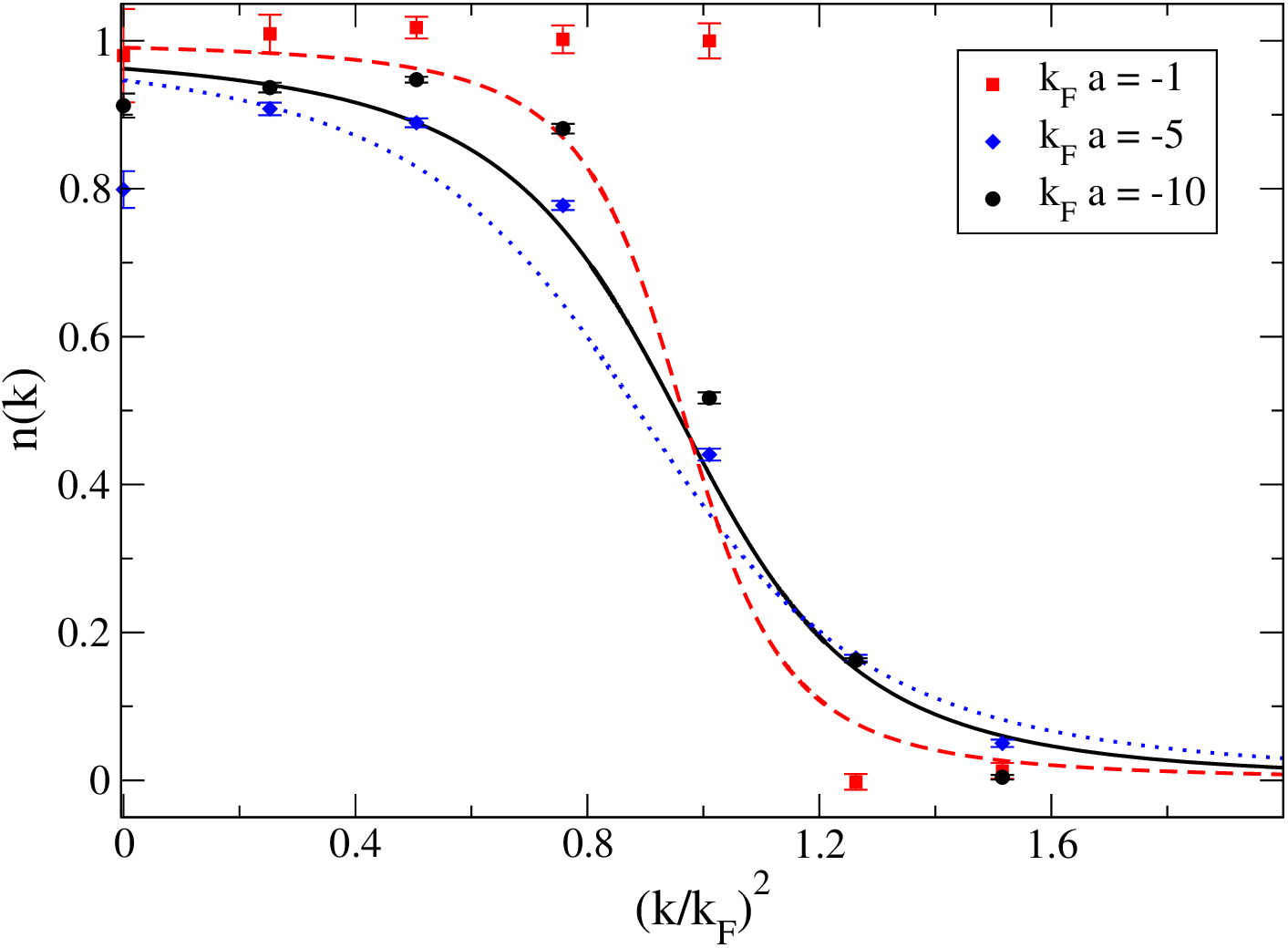}
\caption{(color online) The neutron-matter momentum distribution in QMC versus ${(k/k_F)}^2$ at $k_F a = -1$ (squares), $k_F a = -5$ (diamonds), and  $k_F a = -10$ (circles). Also shown are the continuum BCS results at $k_F a = -1$ (dashed line), $k_F a = -5$ (dotted line), and  $k_F a = -10$ (solid line).}
\label{fig:NMmodi}
\end{center}
\end{figure}

Using the QMC AV4 interaction, we have also calculated distribution functions. In Fig. \ref{fig:NMmodi} we show the momentum distribution at three densities, calculated as the Fourier transform of the one-body density matrix, through:
\begin{equation}
n(k) \equiv \frac{N}{L^3} \left\{ \int d \delta r e^{i {\bf k} \cdot ({\bf r}_{n}' - {\bf r}_{n})} \frac{\Psi_{V}({\bf r}_1, \ldots, {\bf r}_n')}{\Psi_{V}({\bf r}_1, \ldots, {\bf r}_n)} \right\}~,
\end{equation}
where the curly brackets denote a stochastic integration over the angles and we perform the integral over $\delta r = |{\bf r}_{n}' - {\bf r}_{n}|$ on a line using Gaussian quadratures to avoid statistical errors due to the oscillatory radial dependence.

As expected from standard BCS theory, we see that the spread of the momentum distribution around $\mu$ is approximately $2 \Delta$. For $k_F a = -1$ the momentum distribution looks very similar to that of a free Fermi gas. At large $| k_F a|$ (when the gap is approximately half the Fermi energy) this fact implies that there is no clearly defined Fermi surface. We note that
in Fig. \ref{fig:NMmodi} the results for  $k_F a = -5$ seems to be more ``broadened'' than that of $k_F a = -10$, even though as we can see in Fig. \ref{fig:gap-compare} the gap at $k_F a = -5$ is considerably smaller than the one at $k_F a = -10$. This is easily resolved if one looks at the pairing gap not in absolute units (MeV) but divided with the Fermi energy, as shown in Fig. \ref{fig:AFDMClikegap}. The case of $k_F a = -5$ has a bigger gap in units of the Fermi energy, and that is what leads to the observed behavior in the momentum distributions shown in Fig. \ref{fig:NMmodi}.

\begin{figure}[t]
\vspace{0.5cm}
\begin{center}
\includegraphics*[width=0.46\textwidth]{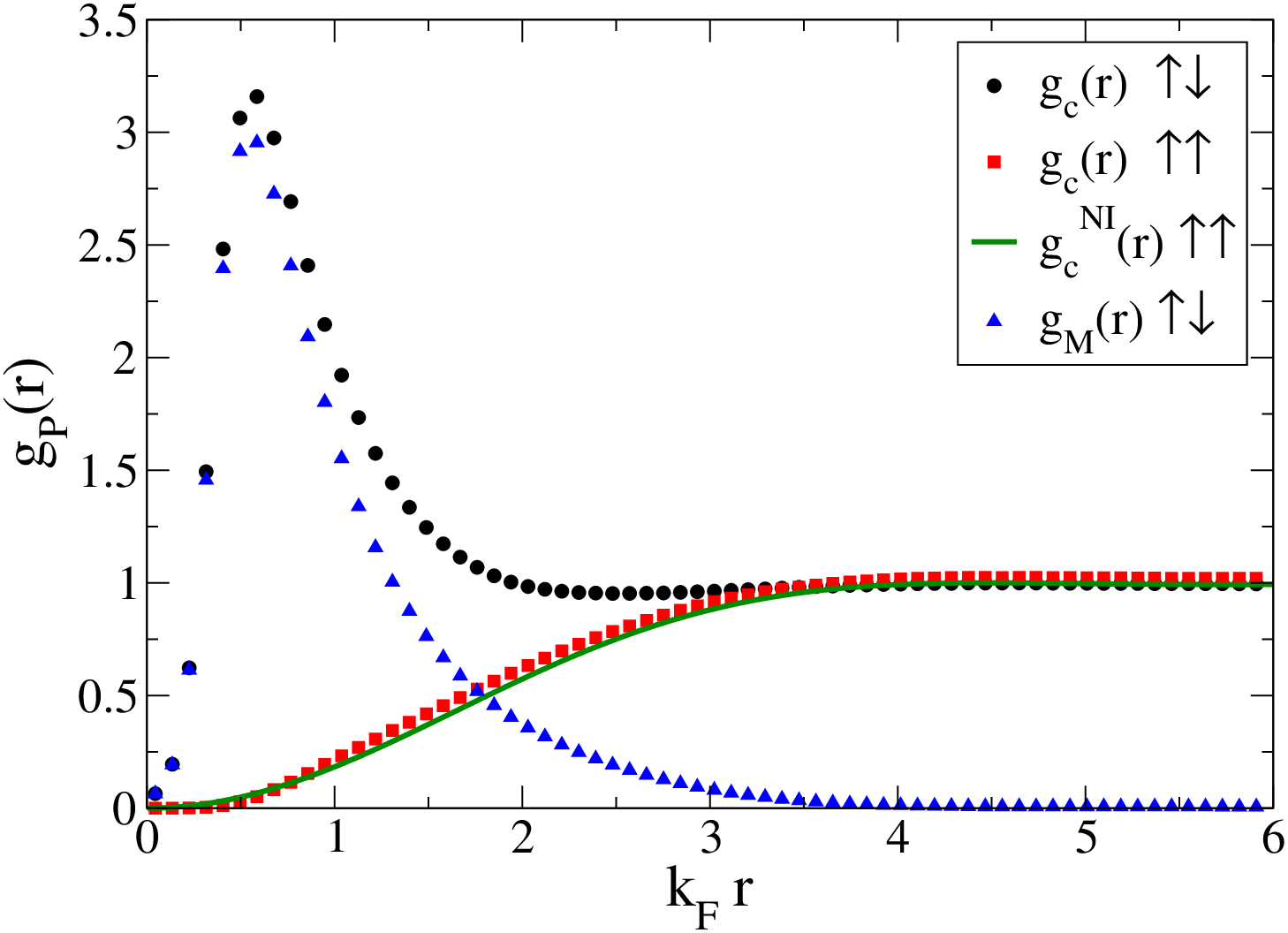}
\caption{(color online) The neutron-matter pair-distribution function in QMC as a function of the distance times the Fermi momentum 
at $k_F a = -10$. The distribution functions given are the $g_c(r)$ for opposite-spin (circles) and 
same-spin pairs (squares), as well as the $g_M(r)$ for opposite-spin pairs (triangles).}
\label{fig:NMpadi}
\end{center}
\end{figure}

We have also computed the pair-distribution functions at $k_F a = -10$ using the AV4 potential, and have plotted them in Fig. \ref{fig:NMpadi}. These are calculated from expectation values of the form:
\begin{equation}
g_P(r) = A \sum\limits_{i<j} \langle \Psi_0 | \delta(r_{ij}-r) O^P_{ij} | \Psi_V \rangle~,
\label{eq:CApadi}
\end{equation}
where we are initially interested in the case in which the operator is simply unity, and the normalization factor $A$ is such that $g_1(r) \equiv g_c(r)$ goes to one at large distances. These pair-distribution functions provide sum rules related to density- and other response functions versus density
and momentum. The solid line in the figure shows the pair-distribution function  of noninteracting (NI) fermions with parallel spins:
\begin{equation}
g_{c}^{NI}(r) = 1 - \frac{9}{(k_F r)^6} \left [ \sin (k_F r) - k_F r \cos (k_F r) \right ]^2~.
\end{equation}
The noninteracting result is very close to the QMC simulation for same-spin particles. On the other hand, the value of the opposite-spin distribution function is very small at short distances, reflecting the repulsive core in the AV18 potential. Since our interaction is more complicated than a simple $s$-wave component, Eq. (\ref{eq:CApadi}) can also be applied to the Majorana exchange operator $P^M$ which was used in Eq. (\ref{vmajorana}). The distribution function for that operator is also shown in Fig. \ref{fig:NMpadi} for the density of interest. It tracks the behavior of the central (unit operator) distribution function for short distances, but then
reduces to approximately half the standard distribution at $k_F r \approx 1.7$.

\section{Conclusions}

To conclude, we have calculated the equation of state and pairing gap of low-density neutron matter with the AV4 interaction from $k_F = 0.054$ to $0.54$ fm$^{-1}$, corresponding to $|k_F a|$ from $1$ to $10$. The calculated equation of state and pairing gap match smoothly with the known analytic results at low densities, and provide important constraints in the strong-coupling regime at large $k_F a$.  
We have also calculated the quasiparticle spectrum,  momentum distribution, and  pair-distribution functions for low-density neutron matter. The low-density equation of state can help constrain Skyrme mean-field models of finite nuclei. The pairing gap for low-density neutron matter is relevant to Skyrme-Hartree-Fock-Bogoliubov calculations \cite{Chamel:2008} of neutron-rich nuclei and to neutron-star physics, since it is expected to influence the behavior of the crust.\cite{Page:2009}

More specifically, a magnetic field in the neutron star crust would have to be approximately $10^{17}$ G to overcome this gap and thus polarize neutron matter; such a value of the magnetic field is not implausible within the context of magnetars.
Similarly, the fact that the magnitude of the gap is not as small as previously expected implies that a new mechanism that makes use of superfluid phonons is competitive to the heat conduction by electrons in magnetized neutron stars.\cite{Aguilera:2009}  

Our results for the gap at the low-density regime, following from a variationally optimized approach that includes  the dominant well-known terms in the Hamiltonian, can function as a benchmark with which other calculations can be compared. Equally important, future experimental tests in cold atoms,
at least in the very low density regime up to $|k_F a| = 2$ appear to be within the reach of possibility.  Similar comparisons may be made with
other observables including the pair-distribution function and momentum distribution.  We believe that these calculations of the equation of state, pairing gap,
and quasiparticle dispersion can be used as constraints of nuclear density functionals.

\begin{acknowledgments}
The authors thank S. Gandolfi, A. Schwenk, and S. Reddy for useful discussions. 
The work of A.G. and J.C. was supported by the UNEDF SciDAC Collaboration
under DOE Grant No. DE-FC02-07ER41457, by the Nuclear Physics Office of the U.S. Department 
of Energy under Contract No. DE-AC52-06NA25396, and by the LDRD program at Los Alamos National Laboratory.  
Computing resources were provided at LANL through the Institutional Computing Program and at NERSC. 
The work of A.G. was supported in part by DOE Grant No. DE-FG02-97ER41014 and by NSF Grant Nos. PHY03-55014 and PHY07-01611.
\end{acknowledgments}




\begin{thebibliography}{99}

\bibitem{Brown:2000} B. A. Brown, Phys. Rev. Lett. {\bf 85}, 5296 (2000).

\bibitem{Stone:2003} J. R. Stone, J. C. Miller, R. Koncewicz, P. D. Stevenson, and M. R. Strayer, Phys. Rev. C {\bf 68}, 034324 (2003).

\bibitem{Chamel:2008}  N. Chamel, S. Goriely, and J.M. Pearson, Nucl. Phys. {\bf A812}, 72 (2008).

\bibitem{Brown:2009}  E. F. Brown and A. Cumming, Astrophys. J. {\bf 698}, 1020 (2009).

\bibitem{Steiner:2009}  A. W. Steiner and S. Reddy, Phys. Rev. C {\bf 79}, 015802 (2009).

\bibitem{Page:2009} D. Page, J. M. Lattimer, M. Prakash, and A. W. Steiner, Astrophys. J. {\bf 707}, 1131 (2009).

\bibitem{Aguilera:2009} D. N. Aguilera and V. Cirigliano and J. A. Pons and S. Reddy and R. Sharma, Phys. Rev. Lett. {\bf 102}, 091101 (2009).

\bibitem{Friedman:1981} B. Friedman and V. R. Pandharipande, Nucl. Phys. {\bf A361}, 502 (1981).

\bibitem{Akmal:1998} A. Akmal, V. R. Pandharipande, and D. G. Ravenhall, Phys. Rev. C {\bf 58}, 1804 (1998).

\bibitem{Carlson:Morales:2003} J. Carlson, J. Morales, Jr., V. R. Pandharipande, and D. G. Ravenhall, Phys. Rev. C {\bf 68}, 025802 (2003).

\bibitem{Schwenk:2005} A. Schwenk and C. J. Pethick, Phys. Rev. Lett. {\bf 95}, 160401 (2005).

\bibitem{Margueron:2007b} J. Margueron, E. van Dalen and C. Fuchs, Phys. Rev. C {\bf 76}, 034309 (2007).

\bibitem{Baldo:2008} M. Baldo and C. Maieron, Phys. Rev. C {\bf 77}, 015801 (2008).

\bibitem{Epelbaum:2008b} E. Epelbaum, H. Krebs, D. Lee, and U. -G. Meissner, Eur. Phys. J. A {\bf 40}, 199 (2009).

\bibitem{Epelbaum:2008a} E. Epelbaum, H.-W. Hammer, and U. -G. Meissner, Rev. Mod. Phys. {\bf 81}, 1773 (2009).

\bibitem{Hebeler:2009} K. Hebeler and A. Schwenk, arXiv:0911.0483 (2009).

\bibitem{Rios:2009} A. Rios, A. Polls, and I. Vida\~na, Phys. Rev. C {\bf 79}, 025802 (2009).

\bibitem{Lombardo:2001} U. Lombardo and H.-J. Schulze, {\it Lecture Notes in Physics} (Springer-Verlag, Berlin, 2001), Vol. 578, p. 30.

\bibitem{Dean:2003} D. J. Dean and M. Hjorth-Jensen, Rev. Mod. Phys. {\bf 75}, 607 (2003).

\bibitem{Chen:1993} J. M. C. Chen, J. W. Clark, R. D. Dav\'e, and V. V. Khodel, Nucl. Phys. {\bf A555}, 59 (1993).

\bibitem{Wambach:1993} J. Wambach, T. L. Ainsworth, and D. Pines, Nucl. Phys. {\bf A555}, 128 (1993).

\bibitem{Schulze:1996} H.-J. Schulze, J. Cugnon, A. Lejeune, M. Baldo, and U. Lombardo, Phys. Lett. {\bf B375}, 1 (1996).

\bibitem{Schwenk:2003} A. Schwenk, B. Friman, and G. E. Brown, Nucl. Phys. {\bf A713}, 191 (2003).

\bibitem{Muether:2005}   H. M\"uther and W. H. Dickhoff, Phys. Rev. C {\bf 72}, 054313 (2005).

\bibitem{Fabrocini:2005} A. Fabrocini, S. Fantoni, A. Y. Illarionov, and K. E. Schmidt, Phys. Rev. Lett. {\bf 95}, 192501 (2005).

\bibitem{Cao:2006}   L. G. Cao, U. Lombardo, and P. Schuck, Phys. Rev. C {\bf 74}, 064301 (2006).

\bibitem{Margueron:2008}   J. Margueron, H. Sagawa, and K. Hagino, Phys. Rev. C {\bf 77}, 054309 (2008).

\bibitem{Gandolfi:2008}  S. Gandolfi, A. Yu. Illarionov, S. Fantoni, F. Pederiva, and K. E. Schmidt, Phys. Rev. Lett. {\bf 101}, 132501 (2008).

\bibitem{Abe:2009} T. Abe and R. Seki, Phys. Rev. C {\bf 79}, 054002 (2009).

\bibitem{Gandolfi:2009}  S. Gandolfi, A. Yu. Illarionov, F. Pederiva, K. E. Schmidt, and S. Fantoni, Phys. Rev. C {\bf 80}, 045802 (2009).

\bibitem{Marcelis:2008} B. Marcelis, B. Verhaar, and S. Kokkelmans, Phys. Rev. Lett. {\bf 100}, 153201 (2008).

\bibitem{Gezerlis:2008} A. Gezerlis and J. Carlson, Phys. Rev. C {\bf 77}, 032801(R) (2008).

\bibitem{Wiringa:1995} R. B. Wiringa, V. G. J. Stoks, and R. Schiavilla, Phys. Rev. C {\bf 51}, 38 (1995).

\bibitem{Lee:1957} T. D. Lee and C. N. Yang, Phys. Rev. {\bf 105}, 1119 (1957).

\bibitem{Gorkov:1961} L. P. Gorkov and T. K. Melik-Barkhudarov, JETP, {\bf 40}, 1452 (1961) [Soviet Phys. JETP {\bf 13}, 1018 (1961)].

\bibitem{Schulze:2001}  H. -J. Schulze, A. Polls, and A. Ramos, Phys. Rev. C {\bf 63}, 044310 (2001).

\bibitem{Khodel:1996}  V. A. Khodel, V. V. Khodel, and J. W. Clark, Nucl. Phys. {\bf A598}, 390 (1996).

\bibitem{Gezerlis:2009} A. Gezerlis, Ph.D. thesis, University of Illinois at Urbana-Champaign, 2009.

\bibitem{Wiringa:2002} R. B. Wiringa and S. C. Pieper, Phys. Rev. Lett. {\bf 89}, 182501 (2002).

\bibitem{Blatt:1952} J. M. Blatt and V. F. Weisskopf, {\it Theoretical Nuclear Physics}, (John Wiley \& Sons: New York, 1952).

\bibitem{Carlson:2003} J. Carlson, S. Y. Chang, V. R. Pandharipande, and K. E. Schmidt, Phys. Rev. Lett. {\bf 91}, 050401 (2003).

\bibitem{Pandharipande:1973} V. R. Pandharipande and H. A. Bethe, Phys. Rev. C {\bf 7}, 1312 (1973).

\bibitem{Schmidt:1999} K. E. Schmidt and S. Fantoni, Phys. Lett. {\bf B446}, 99 (1999).

\bibitem{Gandolfi:pc} S. Gandolfi, private communication (2009).

\bibitem{Shin:2008} Y. Shin, C. H. Schunck, A. Schirotzek, and W. Ketterle, Nature {\bf 451}, 689 (2008).

\bibitem{Carlson:2008} J. Carlson and S. Reddy, Phys. Rev. Lett. {\bf 100}, 150403 (2008).

\bibitem{Schirotzek:2008} A. Schirotzek, Y. I. Shin, C. H. Schunck, and W. Ketterle, Phys. Rev. Lett. {\bf 101}, 140403 (2008).

\bibitem{Carlson:2005} J. Carlson and S. Reddy, Phys. Rev. Lett. {\bf 95}, 060401 (2005).

\bibitem{Bulgac:2008} A. Bulgac and Michael McNeil Forbes, Phys. Rev. Lett. {\bf 101}, 215301 (2008).

\end{thebibliography}
\end{document}